\documentclass[twoside]{dis09}
\usepackage[latin1]{inputenc}
\usepackage[dvips]{graphicx,epsfig,color}
\usepackage{wrapfig,rotating}
\usepackage{amssymb,amsmath,array}

\begin{document}
\title{ {\tiny DESY 09-113 \hfill  SFB/CPP-09-076}\\
Variable-Flavor-Number Scheme in Analysis
of Heavy-Quark Electro-Production Data}

\author{S. Alekhin$^1$, J. Bl\"umlein$^2$, S. Klein$^2$, and S. Moch$^2$
%
%
\vspace{.3cm}\\
%
1- Institute for High Energy Physics \\
Pobeda 1, 142281 Protvino, Moscow Region - Russia\\
%
\vspace{.1cm}\\
2- Deutsches Elektronensynchrotron DESY \\
Platanenallee 6, D--15738 Zeuthen - Germany\\
}

\maketitle

\begin{abstract}
{\small
We check the impact of the factorization scheme employed in the 
calculation of the 
heavy-quark deep-inelastic scattering (DIS) electro-production 
on the PDFs determined in the NNLO QCD analysis of the world inclusive 
neutral-current DIS data combined with the 
ones on the neutrino-nucleon DIS di-muon production and 
the fixed-target Drell-Yan process. 
The charm-quark DIS contribution is calculated in the general-mass
variable-flavor-number (GMVFN) scheme: At asymptotically 
large values of the momentum transfer $Q$ it is given by the zero-mass 
4-flavor scheme and at the value of 
$Q$ equal to the charm-quark mass it is smoothly matched with the 
3-flavor scheme using the Buza-Matiounine-Smith-van Neerven 
prescription. 
The PDFs obtained in this variant of the fit are very similar to the ones 
obtained in the fit with a 3-flavor scheme employed.
Our 5-flavor PDFs  derived from the 3-flavor ones using the NNLO 
matching conditions are used to calculate 
the rates of $W^{\pm}/Z$ and $t\bar{t}$ production 
at the Tevatron collider and the LHC at NNLO.    
}
\end{abstract}

\begin{figure}
\centerline{\includegraphics[width=\columnwidth,height=6cm]
{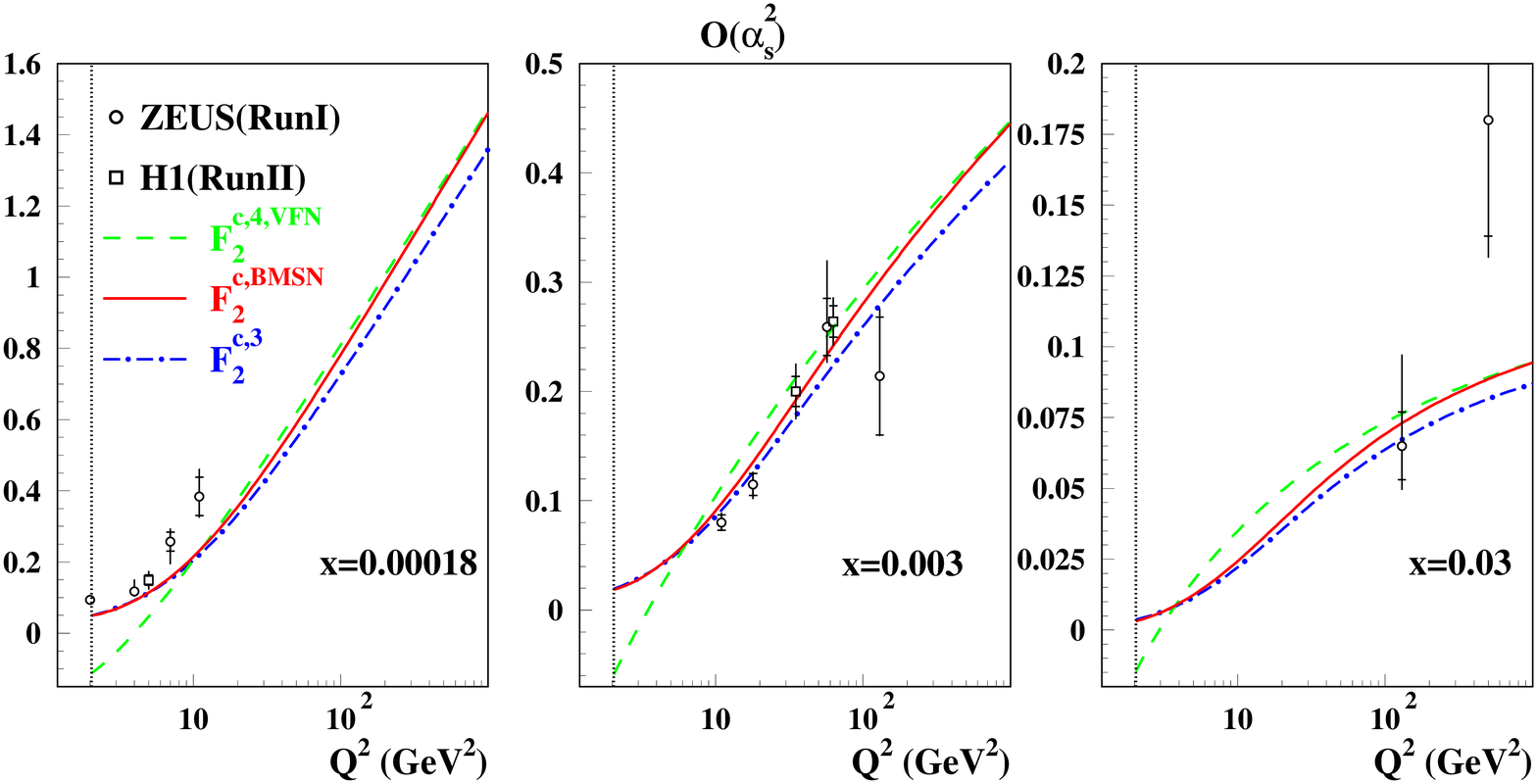}}
\caption{\small The values of $F_2^c(x,Q^2)$ calculated in the different schemes
compared to the data by H1~\cite{H1c} (squares) and 
ZEUS~\cite{ZEUS} (circles) collaborations
(dotted-dashes: the 3-flavor scheme; 
solid lines: the BMSN prescription of the GMVFN scheme,
dashes: the massless 4-flavor scheme).  
}\label{Fig:f2c}
\end{figure}

{\bf 1}. A clear theoretical understanding of the heavy-quark electro-production
is necessary for the phenomenology of lepton-nucleon DIS at HERA since it 
contributes up to 25\% to the DIS inclusive 
cross section at small $x$~\cite{H1c,ZEUS}. This is particularly important for 
the validation of the parton distribution functions (PDFs) 
in the kinematic region 
relevant for the foreseen experiments at the LHC since, despite the impact 
of collider data on the PDFs, DIS remains an unique source of 
information about PDFs at small $x$. Two complementary approaches were suggested
to calculate the DIS heavy-quark contribution. One of them is based on the  
factorization scheme with three light quarks in the initial state. 
In the 3-flavor scheme the heavy-quark production 
is calculated in fixed order of perturbative QCD with complete 
corrections up to the $O(\alpha_s^2)$ and partial corrections 
up to $O(\alpha_s^3)$ 
available~\cite{Witten:1975bh,HQNLO,HQNNLO,Bierenbaum:2008yu}. The 
logarithmic terms in the QCD corrections
rise in the asymptotic region of the momentum transferred
$Q\gg m_h$, where $m_h$ is the heavy quark mass. Therefore they must be 
re-summed for such kinematics~\cite{Shifman:1977yb}.
This leads to the concept of the 
variable-flavor scheme (VFN) with the heavy quarks 
considered in the same way 
as the light-flavor ones. In the VFN scheme 
large-log re-summation suggested in 
Ref.~\cite{Shifman:1977yb} is performed through the QCD evolution 
of the heavy-quark PDFs~\cite{Aivazis:1993pi}. 
The VFN scheme is valid only at asymptotically large values of $Q$ and 
cannot be routinely extrapolated to the low-$Q$ region.
For realistic kinematics it has to be extended to the case of 
a general-mass VFN (GMVFN) scheme, which, at small $Q$, must give 
a smooth matching with the 3-flavor scheme. 
Different variants of the GMVFN scheme
are used in the context of the global fit of PDFs based on 
small-$x$ DIS data~\cite{Thorne:2008xf}.
At the matching point $Q=m_h$ the structure function $F_2^h$
calculated within the GMVFN schemes of Ref.~\cite{Thorne:2008xf} 
is continuous. 
However these schemes do not guarantee a smooth transition of $F_2^h$
between the 3-flavor and the VFN schemes. 
In our study we consider a variant of the GMVFN scheme based on 
the Buza-Matiounine-Smith-van Neerven (BMSN) prescription of 
Ref.~\cite{Buza:1996wv}, which indeed 
provides a smooth behavior of $F_2^h$ at $Q=m_h$,
and perform a phenomenological comparison of this prescription with 
the 3-flavor scheme in the PDFs fit.  

{\bf 2.} The conclusion about the impact of the scheme evidently depends on the 
kinematic coverage and precision of the data used in the fit. 
Figure~\ref{Fig:f2c} shows a comparison of the charm structure function
$F_2^c$ calculated in 
the 3-flavor scheme and the BMSN prescription of the GMVFN scheme 
with the use of the MRST2001 PDFs~\cite{Martin:2002aw} to the 
representative set of the 
recent HERA collider data by H1 and ZEUS~\cite{H1c,ZEUS}. 
The available data on $F_2^c$ are not sensitive to the difference between 
the 3-flavor and the GMVFN scheme in the BMSN prescription. The 
best sensitivity of the data to the scheme choice appears at $x\sim 0.003$, 
however even in this case the errors in the data are quite big as compared to 
the scheme differences. 
This does not look like a specific feature of the 
BMSN prescription since even at maximal values of $Q$ reached by the existing 
experiments the difference 
between the massless 4-flavor scheme and the 3-flavor scheme is smaller 
than the data uncertainties. In particular this happens because 
the NLO corrections of Ref.~\cite{HQNLO} to the heavy-quark 
electro-production greatly 
reduce the need of large-log re-summation~\cite{Gluck:1993dpa}.
With the NNLO corrections of Ref.~\cite{Bierenbaum:2008yu} 
this need should be further reduced.
At small $Q$ the difference between the 3-flavor and BMSN scheme 
is suppressed, similarly to any other GMVFN scheme, which provides 
smooth matching with the former. 
The discrepancy between theory and data at small $Q$ and $x$ 
observed in Figure~\ref{Fig:f2c} cannot be resolved by a scheme change. 
It rather can be cured by the partial NNLO corrections 
of Ref.~\cite{Alekhin:2008hc} due to the 
soft gluon re-summation near threshold. 

\begin{figure}
\centerline{\includegraphics[width=\columnwidth,height=7cm]
{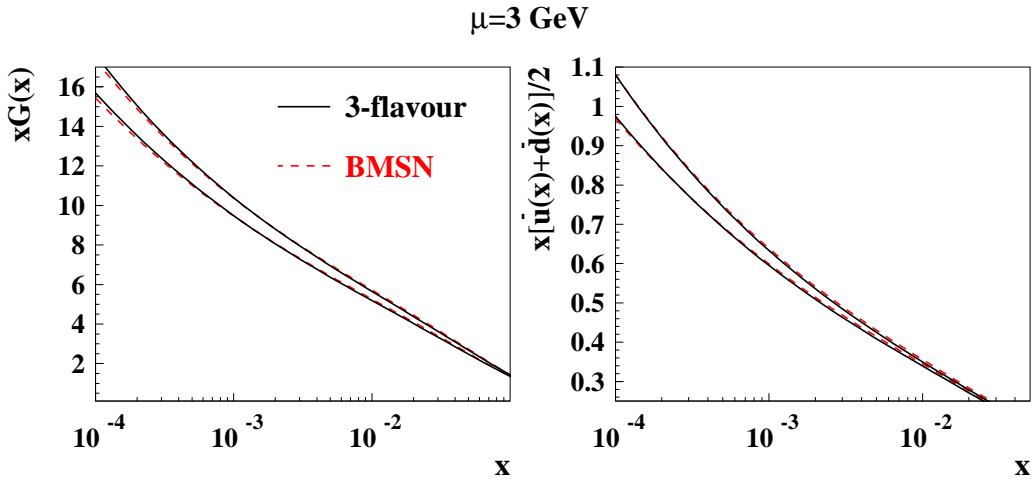}}
\caption{\small The gluon (left panel) and sea (right panel) 
      distributions at the factorization scale of $\mu=3~{\rm GeV}$
       obtained in two variants of the fit. Solid lines: 3-flavor 
      scheme, dashed lines : GMVFN scheme in the BMSN prescription.
}\label{Fig:pdfcomp}
\end{figure}

{\bf 3.} The inclusive DIS data are more sensitive to the choice between 
the VFN and 3-flavor scheme because of higher accuracy. However, 
the PDFs obtained from the fits based on these two variants are 
still similar.
We check this point for the example of a combined fit of PDFs to the inclusive 
DIS data supplemented by the fixed-target Drell-Yan data 
and the DIS di-muon production data, 
which improves the flavor separation of the quark PDFs.
We consider two variants of the fit, with $F_2^c$ calculated in the 
3-flavor scheme or in the BMSN prescription. 
The NNLO corrections to the PDF 
evolution~\cite{Moch:2004pa}, the light-parton coefficient 
functions~\cite{coef}, and 
the NLO corrections of Ref.~\cite{HQNLO} to the heavy-quark electro-production 
coefficient functions are taken into account.
The non-QCD corrections applied and other details of the fit 
can be found elsewhere~\cite{alekhin}. The gluon and sea distributions 
obtained in the two variants of the fit are compared in 
Figure~\ref{Fig:pdfcomp}.
The difference between the two variants of the fit is marginal and is 
in agreement with the 
comparison given in Figure~\ref{Fig:f2c}. In the case of Thorne-Roberts (TR) 
prescription of Ref.~\cite{Thorne:1997uu}
employed for the GMVFN scheme the impact of the scheme choice is somewhat  
bigger~\cite{CooperSarkar:2007ny}. That might be related to a 
kink in $F_2^c$ at $Q=m_c$ for small $x$. For the ACOT 
prescription~\cite{Aivazis:1993pi} and the ACOT($\chi$)
modification of this 
prescription~\cite{Tung:2001mv} employed in the NNLO
analysis of Ref.~\cite{Martin:2009iq}  
this kink is more pronounced than for the case of the TR prescription.
As a result, the heavy-quark contribution to the inclusive 
structure function is overestimated and the light-parton distributions are 
underestimated, correspondingly. 

\begin{figure}
\centerline{\includegraphics[width=\columnwidth,height=6cm]
{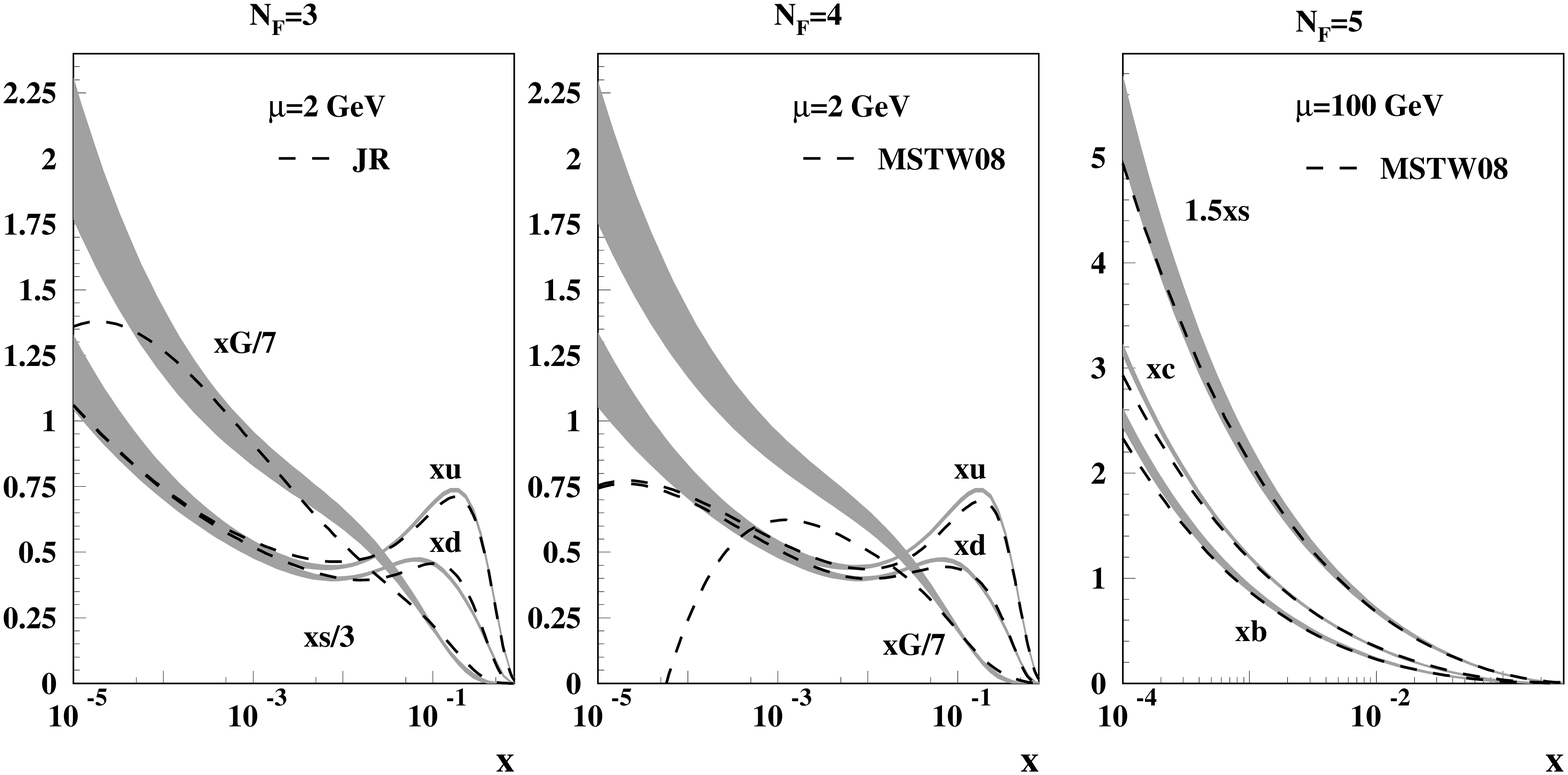}}
\caption{\small The $1\sigma$ band for the representative set of our 
PDFs (shaded area) 
and the central values of other NNLO PDFs sets (dashes). 
Left panel: the 3-flavor $u$-, $d$-quarks and gluons at the scale 
of $\mu=2~{\rm GeV}$
compared to ones of the JR set; central panel: the 4-flavor
$u$-, $d$-quarks and gluons at the scale of $\mu=2~{\rm GeV}$ 
compared to ones of 
the MSTW08 set; right panel: the 5-flavor $s$-, $c$-, and $b$-quarks 
at the scale of $\mu=100~{\rm GeV}$ compared to ones of the MSTW08 set.
}\label{Fig:comp}
\end{figure}

{\bf 4.} Contrary to the case of DIS
for many processes of interest at the energies of the 
Tevatron collider and the LHC the use of a VFN scheme is justified since 
the typical factorization scale is much 
bigger than the $c-$ and $b-$quark masses. 
Moreover, very often such an approach is very efficient since it allows to 
simplify matrix element calculations. For the purposes of collider 
phenomenology applications we generate the NNLO 4-flavor PDFs using 
the matching conditions of Ref.~\cite{Buza:1996wv} with the 3-flavor 
PDFs obtained in our fit as input. Similarly, we use these 4-flavor 
PDFs to produce the 5-flavor ones. Figure~\ref{Fig:comp} shows a comparison of 
our 4-flavor PDFs at the scale of $\mu=2~{\rm GeV}$ with the NNLO MSTW08 
PDFs of Ref.~\cite{Martin:2009iq}. \footnote{At the 
factorization scales close to 
$m_c$ the 4-flavor PDFs are inapplicable. However, we use this set 
for the purpose of comparison
since the 3-flavor PDFs are unavailable for the MSTW08 case.}
At this scale the sea and gluon distributions by MSTW are smaller 
than ours, moreover the MSTW gluon distributions are negative
at $x\lesssim 0.0005$. The difference in the gluon distribution 
of these two sets is sensitive 
to the recent measurement of the structure function $F_L$ by the H1 
collaboration~\cite{FLexp}.
The H1 data are in good agreement to
the predictions of $F_L$ based on our PDFs, however, then go slightly above  
the MSTW08 ones and in that way prefer bigger gluons at small $x$. 
The agreement between our PDFs and the JR PDFs 
of Ref.~\cite{JimenezDelgado:2008hf} obtained in the fit 
performed in the 3-flavor scheme 
\small
\begin{wraptable}{l}{0.55\columnwidth}
\centerline{\begin{tabular}{|c|c|c|c|}   
\hline
$\sqrt{s}~({\rm TeV})$ &  $\sigma(W^\pm)~[nb]$ & $\sigma(Z)~[nb]$ & 
$\sigma(t\bar{t})~[pb]$\\ \hline
$1.96~~(\bar p p)$ & $26.2\pm0.3$  & $7.73\pm0.08$ & $6.91\pm0.17$ \\ \hline
$7~~~~(p p)$ & $98.8\pm1.5$ & $28.6\pm0.5$ & $131.3\pm7.5$\\ \hline
$10~~~~(p p)$ & $145.6\pm2.4$ & $42.7\pm0.7$ & $342\pm15$\\ \hline
$14~~~~(p p)$ & $207.4\pm3.7$ &$61.4\pm1.1$ & $780\pm28$\\ \hline
\end{tabular}}
\caption{\small The integral $W^{\pm} / Z$ and $t\bar{t}$
 production cross sections at the 
energies of 
Tevatron and LHC with the $1\sigma$ uncertainties estimated from the fit 
results in the present analysis.}
\label{tab:wzcs}
\end{wraptable}
\normalsize  
is much better (see Figure~\ref{Fig:comp}). In this case the gluon distributions
differ only at $x\sim 10^{-5}$, in a region not being constrained by data. 
Our 5-flavor $c-$ and $b-$quark distributions are 
larger than the MSTW ones. This is related to the difference in the gluon 
distributions. 
The uncertainties in our PDFs given in Figure~\ref{Fig:comp} are propagated 
from the uncertainties in experimental data with account of their 
correlations. 
We also include the effect of the $c-$ and $b-$ quark mass variation by 
$\pm 0.1$ GeV and $\pm 0.5$ GeV, respectively, into the PDF uncertainties.
For the case of $c-$ and $b-$quark PDFs
the uncertainties in the masses are dominant sources. 
The value of the strong coupling $\alpha_{\rm s}(M_{\rm Z})=0.1135\pm0.0014$
is obtained in our fit, in good agreement with the results of 
Ref.~\cite{JimenezDelgado:2008hf} and earlier results of 
Refs.~\cite{Blumlein:2006be,alekhin}.

{\bf 5.} The rates of the 
 $W^{\pm}/Z$ and $t\bar{t}$ production at the Tevatron collider 
and the LHC obtained with our 5-flavor PDFs are given in Table~1. The 
$W^{\pm}/Z$ rates are calculated with the QCD corrections up to 
NNLO~\cite{Hamberg:1990np}. In the case of the $t\bar{t}$ production 
only partial NNLO corrections of Ref.~\cite{Moch:2008qy}
stemming from the threshold gluon re-summation are taken into account.
The uncertainties in the $W^{\pm}/Z$ rates due to the PDFs vary from 
1\% at the Tevatron energy to 2\% at the maximal LHC energy. Such a level 
of accuracy allows to use these processes as a luminosity monitor.  
Our estimate of the $t\bar{t}$ production rate for Tevatron is 
in agreement to the estimate obtained with the MSTW08 PDFs, while 
at the LHC energies the latter are bigger than ours. As in the 
case of $F_L$ this difference is related to the difference in the gluon 
distributions. Therefore $t\bar{t}$ production at the LHC can be used to 
discriminate between these two PDFs sets. 

{\bf Acknowledgments.}
This work was supported in part by DFG
Sonderforschungsbereich Transregio 9, Computergest\"utzte Theoretische
Teilchenphysik, the RFBR grant 08-02-91024, Studienstiftung des Deutschen 
Volkes, the European Commission MRTN HEPTOOLS under Contract No. 
MRTN-CT-2006-035505.

\begin{footnotesize}

\end{footnotesize}

\end{document}